# Why do drivers and automation disengage the automation? Results from a study among Tesla users

Nordhoff, S. [a], De Winter, J. C. F. [b]

[a] *Department Transport & Planning, Delft University of Technology, The Netherlands*
[b] *Department Cognitive Robotics, Delft University of Technology, The Netherlands*

## ABSTRACT

A better understanding of automation disengagements can impact the safety and efficiency of automated systems. This study investigates the factors contributing to driver- and system-initiated disengagements by analyzing semi-structured interviews with 103 users of Tesla's Autopilot and FSD Beta. Through an examination of the data, main categories and sub-categories of disengagements were identified, which led to the development of a triadic model of automation disengagements. The model treats automation and human operators as equivalent agents. It suggests that human operators disengage automation when they anticipate failure, observe unnatural or unwanted automation behavior (e.g., erratic steering, running red lights), or believe the automation is not suited for certain environments (e.g., inclement weather, non-standard roads). Human operators' negative experiences, such as frustration, feelings of unsafety, and distrust, are also incorporated into the model, as these emotions can be triggered by (anticipated) automation behaviors. The automation, in turn, monitors human operators and may disengage itself if it detects insufficient vigilance or traffic rule violations. Moreover, human operators can be influenced by the reactions of passengers and other road users, leading them to disengage automation if they sense discomfort, anger, or embarrassment due to the system's actions. This research offers insights into the factors contributing to automation disengagements, highlighting not only the concerns of human operators but also the social aspects of the phenomenon. Furthermore, the findings provide information on potential edge cases of automated vehicle technology, which may help to enhance the safety and efficiency of such systems.

# 1. Introduction

Since October 2020, drivers in the United States and Canada have been using Tesla's Full-Self Driving (FSD) Beta feature, an SAE Level 2 system that expands the operational design domain of the standard Autopilot beyond highways to non-highway roads. Tesla cautioned FSD Beta program participants via email to exercise vigilance while using the system, as it may potentially malfunction at the most inopportune moments. Drivers were instructed to maintain their hands on the steering wheel, pay close attention to the road, and avoid complacency (Nordhoff et al., 2023).

Research indicates that Tesla Autopilot users might misuse automation, becoming complacent and engaging in hazardous behaviors such as hands-free and mind-off driving, intentionally manipulating the steering wheel to feign attentiveness, and sleeping while the system is engaged (Nordhoff et al., 2023). The present study focuses on the disengagement (deactivation) of automation, which can be initiated by either the driver (driver-initiated) or the system itself (system-initiated). Deactivation in this study is not limited to immediate responses to environmental occurrences; it may also manifest in a more enduring manner. This enduring deactivation has been referred to as disuse, or the underutilization of automation (Parasuraman & Riley, 1997). More specifically, the term 'disuse' is commonly applied to situations in which the driver chooses not to use automation (i.e., not turning on, or turning the automation off after usage) even though usage would be beneficial for performance and safety. Disengaging automation poses a safety concern: If drivers opt to disengage the automation in situations where it could enhance their performance and road safety, automation's potential benefits will not be realized. Conversely, it could be posited that disengaging automation might be beneficial or even necessary to avert accidents.

The study of human use, misuse, disuse, and abuse of automation has been a subject of research since the 1990s. Scholars have discovered that automation disuse arises from a multitude of factors, including low perceived automation reliability and false alarms, task complexity, risk, learning about automation states, fatigue, a general negative predispositions towards automation and a resistance to innovation, missing functionality, and an aversion to automation's 'bells and whistles' (De Winter et al., 2022; Ferris et al., 2010; Lee, 2006, 2008; Parasuraman & Riley, 1997; Parasuraman et al., 2008; Reagan et al., 2018). These studies propose that users are more likely to engage automation if they perceive its reliability to be greater than their self-confidence. Conversely, if users' self-confidence surpasses their trust in automation, they are more inclined to disengage it. High workload may also prompt users to disengage reliable, accurate, and trustworthy automation.

Research on automation disengagement of automated vehicles demonstrates that driver-initiated disengagements occur in anticipation of potentially hazardous situations, such as adverse weather conditions, construction zones, poor road infrastructure, the presence of emergency vehicles, or navigating curves. Other reasons for disengaging automation encompass executing lane-changing maneuvers, lack of trust or discomfort, and other road users' (reckless) driving behavior (Boggs et al., 2020; Dixit et al., 2016; Favarò et al., 2018; Kim et al., 2022; Lee, 2006; Lv et al., 2017; Wilson et al., 2020). System-initiated disengagements can arise due to missing system functionality (Gershon et al., 2021), such as failures in detection technology, communication, sensor readings, map calibration, or hardware (Dixit et al., 2016).

Understanding the reasons for disengaging automated driving systems (i.e., switched off, or not switched on) is essential to ensure their use in situations where they offer increased safety and efficiency compared to human drivers. Knowledge about disengaging automation can also provide valuable insight into 'edge case' scenarios, namely situations in which human drivers disengage the automated driving system because its operational boundaries have been exceeded. Addressing these edge cases is pivotal to fully realizing the benefits of automation and minimizing its risks (Ryerson et al., 2020), and contributes to the widespread acceptance of automated cars.

There is a scarcity of research on the factors underlying driver- and system-initiated disengagements of Autopilot and FSD Beta. Previous studies have examined various factors contributing to disengagements, such as the impact of automation capability on trust and reliance, as well as the influence of operator confidence, risk perception, and learning. However, previous research has often studied these factors in isolation. Most of the research reporting disengagements of partially automated driving relied on technical vehicle data (Alambeigi et al., 2020; Favarò et al., 2018) rather than in-depth qualitative accounts from the users of these systems. The psychological aspects of disengaging Autopilot and FSD Beta are not well documented. Our study adopts a more comprehensive approach by examining the reasons behind disengagements of complex real-life partially automated driving systems: Tesla's Autopilot and FSD Beta.

## 2. Method
### 2.1. Recruitment

We conducted semi-structured interviews with drivers of partially automated vehicles equipped with standard Autopilot and FSD Beta. The study was approved by the Human Research Ethics Committee of Delft University of Technology.

We recruited users of standard Autopilot and FSD Beta through specialized online communities and forums (i.e., Discord, Facebook, Twitter, Reddit, YouTube, Instagram, Tesla Motors Club, and Tesla Motors Forum). As FSD Beta was only available to drivers in North America and Canada during the study, we focused our recruitment efforts on these regions. Ownership of a Tesla was subjectively evaluated using self-reported data regarding access to Autopilot and FSD Beta.

## 2.2. Procedure

The interviews were conducted online via Zoom, with both audio and visual recorded. The interviews followed a pre-defined protocol consisting of open-ended and closed-ended questions. To reduce the subjectivity of interview research, an interview protocol was created on Qualtrics, and a link to the questions was sent through the chat function of Zoom at the beginning of the interview. This allowed respondents to view the questions directly, enabling them to advance to the next question independently. The researcher listened to respondents' answers to minimize influencing them during the interview. Respondents were encouraged to skip questions already answered. As the questions were standardized and logically ordered, the researcher's intervention was minimal.

The interview protocol comprised two main parts. Initially, respondents provided informed consent to participate in the study. The first part consisted mostly of open-ended questions, while the second part primarily involved closed-ended questions about respondents' socio-demographic profile, travel behavior (e.g., age, gender, education, frequency of Autopilot and FSD Beta use), and general attitudes towards traffic safety.

The appendix presents an overview of the questions asked in the first part of the interview. The present paper only analyzed the questions Q28, "Do you disengage Autopilot and FSD Beta? Why/why not?", and Q29, "Does Autopilot and FSD Beta disengage? When/in which situations?". Respondents were asked to answer each question separately for Autopilot and FSD Beta, reflecting on any differences between the systems. The questions Q1, Q4, Q24–Q26, and Q30–Q35 were addressed in our previous study (Nordhoff et al., 2023).

## 2.3. Data analysis

Data analysis was performed by the first author of the present study in four steps:

1. Interviews were recorded via Zoom and transcribed verbatim using Microsoft Teams transcription software. Transcripts were compared with audio files and corrected as necessary.

2. Atlas.ti version 22.0.2 was used to create main categories and sub-categories for data analysis. These categories were developed using the categorization of disengagement causes from Boggs et al. (2020), which classifies them in terms of control discrepancies, environmental and other road users, hardware, software, perception, and planning discrepancies, and operator takeover. Additional sub-categories were formed following principles of inductive category development by Mayring (2000) for those not fitting within the proposed classification scheme. The main categories and sub-categories were used to develop a triadic model of driver- and system-initiated disengagements.
3. All sub-categories reported in Table 2 are mentioned at least five times by respondents.
4. Illustrative quotes were chosen to convey the meaning of each sub-category. Multiple mentions of a sub-category by a respondent were not discarded but combined with other mentions of the sub-category by the respondent. As a result, some quotes represent collections of sentences mentioned by the same respondents at different points during the interview. Filler words and repetitions (e.g., "you know", "like", "uhhm") were omitted from the quotes.

## 3. Results

The majority of respondents were male (91%), with an average age of 42 years, highly educated (52% held a Bachelor or Master degree), and occupied positions as engineers (30%), managers (8%), or were retired (7%). They predominantly resided in California (20%), Colorado (8%), and Florida (7%). Eighty-two percent of respondents utilized both standard Autopilot and FSD Beta, while 18% reported having access only to standard Autopilot.

The analysis of interview data led to the development of a triadic model of automation disengagement, as depicted in Figure 1. This model was derived from the main categories and sub-categories presented in Table 1.

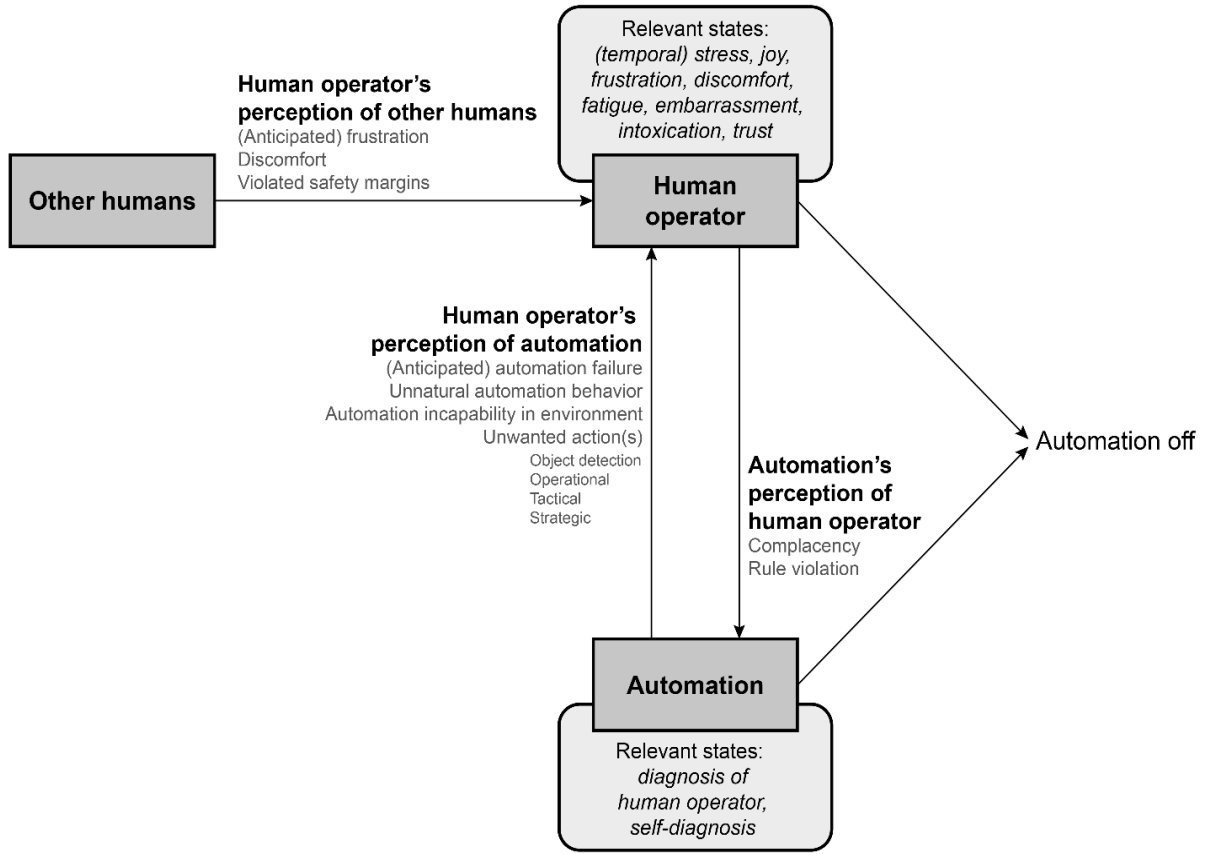

*Figure 1.* Triadic model of automation disengagements.

The model considers the automation and human operator as equivalent agents since the decision to disengage can be initiated by either party. Both agents assess each other's performance and capabilities.

The model identifies both relatively permanent and more transient states of the automation and human operator. More permanent reasons to disengage the automation by human operators is the preference for manual control to enjoy driving. Certain (anticipated) automation behaviors can result in negative transient states experienced by the human operator, such as frustration, stress, and embarrassment.

The human operator may decide to disengage the automation in anticipation of automation failure in environments that exceed the system's capabilities or when the automation exhibits unnatural or unwanted behavior.

Moreover, the human operator can apply a 'theory of mind' to deduce whether other humans might become uncomfortable or frustrated by the automation's actions.

Similarly, the automation may perceive the human operator as insufficiently vigilant (based on the hands-on-wheel sensors) or violating speed limit rules, or it may assess its own capabilities as inadequate in the given environment, leading to self-disengagement.

*Table 1.* Factors of driver- and system-initiated disengagements (main category, sub-category, meaning, type of disengagement (i.e., driver, system), type of system (Autopilot, FSD Beta)

| Main category | Sub-category | Meaning | Type of disengagement | | Type of system | |
|---|---|---|---|---|---|---|
| | | | Driver-initiated | System-Initiated | Autopilot | FSD Beta |
| **Human operator's relevant states** | | | | | | |
| | **Permanent** | | | | | |
| | Fatigue and intoxication | Fatigue and intoxication | X | | | X |
| | Travel trip constraints | Travel time, distance, and purpose constraints, with drivers initiating disengagement due to facing time constraints (in a hurry), travel distance constraints (distance too short), or travel trip constraints on the first and last part of the trip, with the system not being able to pull in and out of the driveway and initiate a parking maneuver | X | | X | X |
| | Preference for manual control to enjoy driving | Preference for manual control to enjoy driving | X | | X | X |
| | **Transient** | | | | | |
| | Frustration and (temporal) stress | Frustration and stress associated with behavior of automation | X | | | X |
| | Embarrassment | Feeling of embarrassment towards other road users due to behavior of automation | X | | | X |
| **Human operator's perception of automation** | | | | | | |
| | Software release | New software releases and bugs associated with software release, with drivers reporting they engage in short-term usage behavior of the automation to test and experience its limits or temporarily stop using it, respectively | X | X | | X |
| | (Anticipated) automation failure | To avoid potential collision before automation can perform the maneuver in situations in which operator capability in their own skills exceeds perceived system capability to handle the situation safely due to underlying feelings of operator discomfort, low perceived safety, and lack of trust, contributing to high frequency of disengagements | X | | | X |
| | Unnatural automation behavior | To correct for unnatural and unhuman automation behavior | X | | | X |
| | Random disengagements | Random disengagements without drivers knowing the reasons underlying disengagements | | X | X | X |

|  | False positives | Disengagements despite drivers supervising automation |  | X | X |  |
|--|--|--|--|--|--|--|
|  | **Unwanted action(s)** | | | | | |
|  | **Operational** | | | | | |
|  | Harsh deceleration | Harsh and sudden deceleration, commonly known as 'phantom braking', associated with shift of Tesla's sensor suite from radar to vision-only, potentially increasing incidents of rear-end collisions and contributing to temporary disuse of automation |  | X | X | X |
|  | Erratic steering wheel movements | Erratic, jerky steering wheel movements of automation | X |  |  | X |
|  | Steering into adjacent traffic | Steering into adjacent traffic | X |  |  | X |
|  | **Strategic** | | | | | |
|  | Route unfamiliarity | On unfamiliar roads due to driver's lack of trust in automation to handle these areas safely given lack of experience in these areas | X |  | X | X |
|  | Taking the wrong route | Taking the wrong route, possibly because of misalignment between mapping and navigation data | X |  | X | X |
|  | **Tactical** | | | | | |
|  | Longitudinal control | Inappropriate control of automation over braking and acceleration, with drivers adjusting operating speed to flow of traffic, increasing speed to close the gap between Tesla and vehicle in front, overtaking a slower vehicle, and decreasing operating speed for safety, comfort, and efficiency | X |  | X | X |
|  | Undesired lane changes | Unexpected, unnecessary, too aggressive, or conservative lane changes | X |  | X | X |
|  | Misidentifying correct lane | Steering into parking, bike, or bus lane | X |  |  | X |
|  | Creeping | Creeping at intersections and in turning situations | X |  |  | X |
|  | Rolling through stop signs | Rolling through stop signs, representing an illegal yet accepted human behavior | X | X |  | X |
|  | Running red lights / stop signs | Running red lights and stop signs | X | X | X | X |
|  | Turning on red | Turning into the direction of traffic at a traffic light showing a red signal | X |  |  | X |
|  | Unprotected left turns | Turning left at intersection with a green light instead of a green arrow with oncoming traffic having the right of way | X |  |  | X |
|  | Other turning situations | Other turning situations, such as protected left-hand and right-hand and U-turns | X |  |  | X |
| **Human operator's perception of other humans** | | | | | | |
|  | Discomfort | Passenger discomfort with perceived lack of control, or unpredictable and erratic system behavior | X |  | X | X |
|  | Road rage | To avoid confusion and rage of other road users | X |  |  | X |
|  | Reckless behavior | To respond to reckless driving behavior of other drivers, following too close from behind, or swerving into the lane of the vehicle | X |  |  | X |
| **Automation's perception of the human operator** | | | | | | |
| | Complacency | Drivers failing to monitor the system | | X | X | |
| | Inappropriate amount of torque applied to the steering wheel | Drivers accidentally disengaging automation by applying inappropriate amount of torque to the steering wheel | X | X | X | |
| | Speed limit violation | Drivers exceeding the speed limit for reasons of safety, efficiency, lack of knowledge, or complacency | X | X | X | X |
| **Automation incapability environment** | | | | | | |
| Weather | Inclement weather conditions | Inclement weather and poor visibility conditions | X | X | X | X |
| Road infrastructure and design | Non-standard roads | On roads with inconsistent and missing lane markings and with non-standard lane width | X | X | X | X |
| | Curves, hills | In curves and hills | X | X | X | X |
| | Objects and events | To respond to stationary (e.g., potholes, road debris, speed bumps, bushes/trees, buildings, parked cars, construction, railway crossings, gates), and non-stationary objects (e.g., vulnerable road users, emergency vehicles, and other vehicles) in the environment due to automation not being capable of identifying and responding to these objects, with drivers consequently swerving around them, or lane-centering of system | X | X | X | X |
| | Intersections | In intersections, including roundabouts | X | X | X | X |
| | Discontinuities in road design | On-ramp and off-ramp, lane merging, and splitting situations | X | X | X | |
| | Complex, heavy traffic | In heavy, complex traffic situations | X | | | X |

## 3.1. Human operator's relevant states

As shown by our model in Figure 1, human operators initiate the disengagement due to permanent and transient operator states. Permanent human operator states include fatigue and intoxication, travel trip constraints, and the preference for manual control to enjoy driving. Transient human operator states include frustration and (temporal) stress as well as embarrassment.

### 3.1.1. Fatigue and intoxication

This sub-theme covered driver-initiated disengagements resulting from drivers being physically or mentally impaired by factors such as fatigue and intoxication.

> *"I stopped using FSD Beta when I'm fatigued or something, and I just don't feel like paying that close attention."* (R012)

> *"If I'm tired at the end of the day, I'll actually just turn off the Beta and drive myself because I know I need to pay attention, and I don't want to have to go* *through this stress of figuring out 'OK, is it going to slow down in the middle of the intersection before proceeding?'"* (R017)

*"With the Beta, I didn't use it on that trip because I was tired, and it was a long trip, and Beta is one of those things you have to be on top of it. You can't use it when you're impaired in any way, when you're super tired, when you've had any alcohol."* (R096)

### 3.1.2. Travel trip constraints

Human operators reported to disengage FSD Beta when they faced travel time constraints, wanted to travel short distances, and at the beginning and end of a trip.

*"If I'm in a little bit of a hurry, I'll take over myself because it is probably slower. It's more careful and cautious than I would in some ways so if I'm trying to save time, I'll take over myself."* (R021)

*"When I take my lunch – it's just a short 15-minute break – it's sort of 'OK, I'm just gonna rush really quick right around the corner, grab a sandwich, and then rush right back to my apartment', and it's such a short distance that it doesn't really warrant using it. So, that would be about the only time that I don't use it."* (R049)

*"The second time that I will disable it will be if I need to get to my destination on a less relaxed timeline, because FSD is great for obeying all the laws that it's supposed to, and there are times where you're running late, you're gonna be in a little bit of a hurry."* (R061)

*"I usually can't engage Beta from my driveway. I can't tell it 'Leave my driveway' or 'Back into my drive'. I would love to teach the car 'Hey, this is how I backed into my driveway. Just repeat that.' So, once I get on my main road, I will usually engage it."* (R071)

*"I turn off FSD Beta when I have to make a tight turn in and out of parking lots because the camera system is limited in what it can see, and it will nose*

*out and then stop, or it will nose out and keep going, and you don't want it to."* (R099)

### 3.1.3. Preference for manual control to enjoy driving

The preference for manual control to enjoy driving was mentioned as one of the reasons leading to disengage automation. This sub-category represents a permanent human operator state as shown by our model in Figure 1.

> *"I stopped using Autopilot when I wanna do the driving because I'm enjoying doing the driving."* (R012)

> *"The other times that I will disengage Autopilot is if I want to have a little bit of fun. The car will never use its full acceleration capability when it's in Autopilot or full self-driving, and so if I'm the lead person at a stoplight, and then I'm watching the surroundings, and see that there's nobody that's going to run a red, I'll turn off Autopilot so that I can be more aggressive, and then turn it back on once we're into our normal speeds."* (R037)

> *"Almost the only time I ever turn it off is if I'm doing a joyride, or in the mountains, and really wanna experience the driving experience. Similar to some of the same motivations I use whenever I do auto cross or rally cross type of stuff as a hobby on closed tracks."* (R065)

> *"Sometimes I'll turn it off because I wanna feel driving for an afternoon on a curvy highway road or something."* (R085)

### 3.1.4. Frustration and (temporal) stress

This sub-theme covered driver-initiated disengagements resulting from frustration and driver's stress with the behavior of FSD Beta.

> *"FSD Beta – I think it's less stressful to not have it enabled."* (R003)

> *"The only thing that impacts whether or not I use it is 'Do I want to not have to worry about driving on a freeway?' I'll use Autopilot if I don't really care.*

*FSD Beta – it's not an unsafe thing that stops me from using it. It's annoying to use it. I don't want to babysit the car, so I'd rather just drive myself."* (R055)

*"With FSD Beta, I see a lot of disengagement on the trip, I just disengage it, and drive away because at some point it becomes dangerous, and it's easier for me. I get frustrated. Like 'What are you? Why? Why are you doing this?' 'What's the point of doing this?' It's just unnecessary."* (R058)

*"There is something in that drive that just gets me to say 'I'm done'. I'm gonna make sure that I don't get frustrated, and that it doesn't make a mistake."* (R061)

*"With the Beta, I was frustrated, and we had gone through a couple updates, and nothing was getting better in my area. I just get angry, and I'm like 'You know what? This is not even usable.' So, I just stopped using it for a while."* (R096)

### 3.1.5. Embarrassment

This sub-theme addressed driver-initiated disengagements resulting from the driver's feeling of embarrassment towards other road users due to the behavior of the FSD Beta.

*"I feel shame and embarrassment because it's like 'Oh, if I turned on my blinker and they gave me room, but I didn't actually need to change lanes.' Now they are confused, and I feel bad because I'm the worst driver on the road. I wanted a bumper sticker like 'It's not my fault. I'm so sorry. It's a student driver.'"* (R007)

*"Just a couple days ago, I was using it, and there was a woman and her dog standing on the corner of the road and I thought 'I'm gonna turn off Full Self-Driving Beta because even though it's probably not gonna hit this woman and her dog, it might turn really stupidly, and then this woman's gonna be like 'Wow, why does she make such a horrible turn?' Would be embarrassed."* (R059)

> *"When it came up into that intersection, all of a sudden it jerks to the left for no reason. It just literally jerks you into the next lane with no notice, no reason, and that's the scary thing. Especially if there was somebody there, it would be like 'Was this guy drunk? Is he texting?'"* (R069)

> *"If I allowed to do auto lane change, sometimes I will interject because it'll be like 'I wanna change back lanes' but then I know 5 seconds later, it's gonna come back out of that driving lane again to pass another vehicle. So sometimes I may turn off the auto lane change feature. Just keep it more consistent versus me look like a crazy driver just zipping back and forth."* (R071)

> *"Most often, when it continues to go, it already stopped but the person is thanking me for stopping, and in that time that the person thanks me, the car decides 'Well, if you're not gonna go, I'm just going to go instead.' It confuses the person. It makes me look bad. It's just bad."* (R074)

## 3.2. Human operator's perception of automation

In addition to disengaging automation due to permanent and transient operator states, human operators may also choose to disengage automation due to new software releases, in anticipation of automation failures in environments exceeding the capabilities of the automation, or when the automation behaves in an unnatural or unwanted way.

### 3.2.1. Software releases

This sub-theme addressed driver-initiated disengagements caused by new software releases, with drivers reporting engaging FSD Beta after a new software update has been released to test and experience the limits of the system, or temporarily disengaging FSD Beta due to software bugs associated with the software release.

> *"FSD Beta – so with each update I like to test the new update on the same sections of road, and if it does something really bad a few times in one section of the road, I won't use it for that whole update, but then I'll try it again in the next update, and it might improve so that I don't keep using it."* (R043)

*"If there is a bad update, I probably just wouldn't use it at all it. If they released an update that just completely broke something, I probably just would stop using it until the next update."* (R063)

*"If the FSD Beta software is just a bad release at least for my area, I just stopped using it most of the time because it's pretty much unsafe."* (R078)

*"There was one particular update they released that had a lot of bugs, and that got people kind of like 'Ohh, no, I don't want to use it', but they fixed it within 4 hours. So, I think the only way that would make me not to want to use FSD is that if I had a buggy software that they released."* (R091)

*"So, for me I don't use it every single day because without another update, it pretty much acts the same way. So, unless I don't have an update, there's no sense for me to keep driving the same area. I'm not gonna get a different result."* (R094)

### 3.2.2. Anticipated automation failure

This sub-theme addressed driver-initiated disengagements in anticipation of upcoming system failure, with drivers reporting a high number of driver-initiated disengagements.

*"I think the furthest I've made it so far is maybe a kilometer before I had to do something."* (R045)

*"I don't think I could have FSD Beta running for over 10 minutes straight without me needing to disengage."* (R055)

*"There is some point where we were going to lunch, my programmer and I, and I took it the two miles into town. It was terrible. All was terrible. It was so bad it would disengage every quarter mile, which is terrible."* (R067)

Drivers reported taking over control before the car executes the maneuver because of discomfort, a lack of trust in the system's capabilities, and a low feeling of safety. Some of these disengagements might have been unnecessary as it is not clear whether the system would have caused a crash if the driver had not intervened.

> *"Sometimes you have traffic coming towards you and there's a hole in traffic, and you can make a left turn. So, the question is how? Is the car going to accelerate through that hole in a way I'm comfortable with, and sometimes I'm just not willing to find out. I'm gonna put my foot on the accelerator because finding out the answer to this question is very significant. It's an accident."* (R012)

> *"You immediately have to intervene now. You don't know if it would have turned itself because you intervened. It's likely that it would have turned. I mean, even Tom was riding with me at that point. I was driving, and he said 'Yeah, it would have turned, but it looks scary.'"* (R067)

> *"I think to myself 'Holy shit, if I didn't take over in less than a half second, I am going to have a head-on collision going 40 miles an hour with another car going 40 miles an hour. I think that it would have corrected itself but then at the same time it's the repercussions, if it didn't correct itself ..."* (R068)

> *"I think the first times I showed it to friends and family, it tried to kill us a couple times, and that's part of the experience. If I just completely let go that wheel, it probably would have done something that would have caused an accident."* (R074)

> *"The system is going haywire and it's like 'Hey, you know, I'm Terminator. I'm gonna do my own thing', and then I feel unsafe forward to the point where I disengage, I turn off FSD, and I'm like 'I don't want to risk anything'."* (R075)

### 3.2.3. Unnatural automation behavior

As shown in our model in Figure 1, disengagements by human operators may occur due to the automation's unnatural or unhuman behavior.

> *"I'm not always sure that's going to do the right thing. Usually it does, but again it doesn't always do it soon or not soon, or at least as early as I would do it myself so there's that divergent between what I would do and what the car is doing. I pretty much always take over at that point."* (R003)

> *"I don't understand if the car is gonna behave the way I would behave or better, and this is probably because in a lot of cases when there's no risk in these scenarios, the car sometimes will do behaviors that are like 'OK, so why did you just sort of come to a stop right here?' If there were other cars around, I'm gonna wanna intervene."* (R012)

> *"When I feel uncomfortable, I take over quickly. It probably would be able to make it through the corners safely. It's just … it doesn't drive like me, which I think is one of the biggest problems, and one of the things that people expect from a full self-driving car is that it would be a comfortable driving experience, but at the moment it does not achieve that."* (R038)

> *"Do I disengage Autopilot, FSD Beta? FSD Beta all the time and that's just because of how I drive. I don't want it to do something that seems inhuman."* (R055)

> *"It will start doing something that at least as a human, I would not do that, and that's typically when I would take over or override it."* (R090)

### 3.2.4. Random disengagements

This sub-theme addressed system-initiated disengagements caused by system error without drivers understanding why the system disengaged.

> *"I had a couple of situations where I was driving on the road, not a regular road, and it just started beeping, and asking me to take control, and I could not tell what exactly it was that it did not like but I just took control for 30 seconds, and then engaged it by the next corner and it was fine."* (R001)

*"FSD Beta disengages more often than not for me, when the system itself is having an error. So, if it says take over immediately and I take the wheel, system will disengage, and say it aborted due to a system error."* (R064)

*"There are times where both of them will disengage for no reason. I'll get the red steering wheel, and it'll say take over immediately, and I don't know why. I work on the overnight shift, so I'll be on the freeway in the middle of the night with no one around me, and the red steering wheel will come on, and it'll tell me you take over immediately, and I'll start slowing down for absolutely no reason. I've had been to do that a couple times where I was at a stoplight, and it just kicked me out and did the same thing. Just said take over!"* (R076)

*"So just two days ago I had an Autopilot software crash. I'm not entirely sure what happened, but it does occasionally happen during this test process. I've never seen FSD Beta crash."* (R094)

### 3.2.5. False positives

Human operators reported that the system disengaged itself despite them supervising the automation.

*"I rest my hand on the wheel while it's dragging, and it will say my hand is not there. It said there is no pressure on the steering wheel so it will disable it. If somebody is not paying attention, you could cause an accident."* (R008)

*"Autopilot itself will give you 'Please keep your hands on the wheel even though your hands are on the wheel.' It gives you that warning regardless, and when it says 'Take over immediately', it will disengage even if you don't have your hands on the wheel. It typically disengages about 530 seconds later even if you refuse to take over."* (R032)

*"There's been a time where I was wearing sunglasses, and it couldn't be sure that I was actually paying attention, and so it gave one of the red steering wheel warnings to take over, and that's essentially the biggest times that it'll disengage if it thinks that you're not paying attention."* (R037)

*"If I have to get yelled at by the camera all the time because it's literally tracking my eyeballs, and it's wrong sometimes. If it sees a phone, it has really good AI, even if you're looking at it, a map or whatever, it just sees the device in your hand, and then you get a strike, right? So, I have two strikes out of five already, and I didn't buy a car to get a strike out."* (R042)

### 3.2.6. Unwanted action(s)

Human operators additionally deactivated automation as a result of undesired automated system actions across operational, strategic, and tactical decision-making dimensions, as depicted in the model presented in Figure 1.

### 3.2.6.1. Operational

**Harsh deceleration**

This sub-theme addressed system-initiated disengagements caused by harsh and sudden deceleration. Drivers associated incidences of harsh braking with the shift of the sensor suite from radar to vision and considered it a safety risk, as it could potentially lead to more incidents of the Tesla being rear-ended. After encountering incidents of harsh braking, drivers were more likely to disengage the system.

*"I was on the Interstate one time in the middle of the day, perfect weather, and it did an automatic emergency brake, and there were some cars behind me, and it almost caused somebody to hit me because of this sudden thing. I checked the reporting, and there was nothing on it. People refer to that as phantom braking. It literally slammed on the brakes, and went from 80 miles an hour down to like 40 miles an hour almost instantly. So, for the next month or so I didn't use it at all."* (R046)

*"When they went to the vision only, I was encountering a lot of phantom braking. There were elements of safety involved. If I'm on the highway, my car just decides to randomly brake hard. If there's someone behind me, that might be an accident. So, I just stopped using it unless there was no one around me. 'Nope. OK, I'm out for a while. I don't need that again.'"* (R056)

*"When it does that phantom braking, then it makes me feel anxious, and I immediately leave Autopilot."* (R059)

*"I was driving in the middle of the desert where you could see ten miles ahead. There's no cars on a two-lane road, and it was just slamming on the brakes. So, at any time I have to be ready. It could just slam on the brakes. That's been my recent experience with FSD Beta."* (R068)

**Erratic steering wheel movements**

Another unwanted automation behavior represents erratic steering wheel movements by FSD Beta, leading to human operators disengaging the automation.

*"If we're approaching a junction, it starts squirming the wheel back and forth. It does it with a lot of force. If it pulls the wheel really hard, you can stop it instantly."* (R051)

*"A lot of times when I set it into FSD, and it just does dumb things like whipping in and out of lanes, I'm just gonna disengage it, and I just drive manually."* (R058)

*"It also increases my anxiety when it does unnecessary wiggles in the lane for unnecessary reasons, thus create higher anxiety because you think there's something I didn't see. Did a cat run out in front of you? You overcorrect, which could cause you to swerve into another car."* (R071)

*"You don't ever know what it's gonna do. I've had it recently. Jerks the wheel back and forth. 'What are you doing? This is not what we do.' So, you slam on the brake, and you take over and drive."* (R084)

*"I only had one hand on the wheel, my elbow was on the armrest, and it swerved. It made this maneuver so strong that it actually bent my wrist down, pulled my arm off the armrest, and I was shocked because I had never had something like that happened now. I saved the video from the dashcam thing to go back and look because I was 'Maybe I missed something. Maybe there*

*was a person that I didn't see.' There was nothing. There was no animal there. There was absolutely no reason for it to do that maneuver."* (R096)

**Steering into adjacent traffic**

The occurrence of FSD Beta steering towards neighboring traffic represents an unintended automated system behavior, which contributes to instances of driver-initiated deactivations of the system.

> *"It was just a straight road. All it needed to do was just drive straight. Stay in the lane lines. Don't hit the car in front of me. Maintain speed. Instead, it tried to swerve left into a vehicle next to me, so I had to disengage."* (R007)

> *"It will take the outside lane until the last second, and it will cut over into the inside lane, which can be an accident, and so you still have to babysit it. It's not a feature yet. It will turn into traffic if you let it."* (R011)

> *"It drove into the wrong lane. It got confused, and then actually drove into the oncoming traffic lane. I've had that happen."* (R042)

> *"Actually, the very first time I used FSD Beta, I turned it on my street. It goes down, and it tries to veer into a car immediately, and I was like 'Ohh no, this isn't good.'"* (R066)

> *"FSD Beta is downright scary. I don't wanna be crude, but it almost makes you piss your pants sometimes when it jerks out of a lane or jerks into a lane. When it does some of that stuff, it's downright scary, and the fact that if you weren't paying attention, then actually took control of the car. You get into an accident, hurt yourself or kill somebody or kill yourself or whatever. There's nothing positive about that."* (R069)

### 3.2.6.2. Strategic

**Route unfamiliarity**

This sub-theme addressed driver-initiated disengagements on unfamiliar roads due to driver's lack of trust in the automation's capability to handle these areas safely given their lack of experience with the automation in these areas, and a desire to be in control.

*"Full-Self-Driving Beta – sometimes when I'm in an area where I'm not familiar with, so I'm not familiar with the roads, and that's when I would disengage it."* (R020)

*"If I'm in a situation where I don't drive very often, then I'm likely to keep FSD Beta on a very short leash. I casually drive into New York City, but once it gets in a crowded situation coming through, or it starts acting a little wonky, I'm gonna take it down, and say 'Hey, I'm not comfortable using it there. It may be able to handle it, but I haven't tested it in this environment enough to feel comfortable doing it.' I intended to not use it in those situations, just because I don't have enough experience."* (R026)

*"FSD Beta – in areas that I drive common, I use it almost all the time. If I were to go into a new city, I most likely won't use it. Maybe if I'm there for example a week, maybe the first few days I won't use it just because I want to see what the area and roads are like and then I use FSD."* (R075)

*So, for Autopilot, I engaged it during interchanges, or maybe an exit, or there's an unfamiliar area, and I wanna stay in control. If I'm not comfortable in an area, I will definitely not let Autopilot take the lead."* (R094)

**Taking the wrong route**

This sub-theme addressed driver-initiated disengagements caused by the system taking the wrong route. These system errors were attributed to a misalignment of mapping and navigation data.

*"It can be driving along, and then be like 'I'm gonna turn left.' Turn left with no blinker or anything. So, let's say it was supposed to turn left, but it didn't, and it continued straight, and then the navigation re-navigates, and just suddenly turned left across multiple lanes."* (R002)

*"It sometimes will do really dumb directions. It's like 'Just turn left here. You don't need to go right, and go around the square and then come back around.'*

*You can't easily necessarily correct that in the moment, so those are instances where you have to disengage."* (R007)

*"I would say that the majority of my disengagements in 10.2 are related to navigation. So right here[1], I was pulling into a sort of a strip mall with a restaurant, and you can see that it takes the turn just fine, but right here[2], we see a very strange road marking where it turns into two lanes without any sort of land marking, and you can see here that it can't visualize it that way. It has its turning right into nothing. There's no parking lot over here, and so I disengage to go to the actual destination that I'm going to."* (R011)

*"One of the more problem areas is that sometimes the map data and information associated with FSD Beta is not the most accurate, and you may have to take over in order to get to the correct location."* (R037)

*"There is one specific corner on the freeway where the map data is wrong, so I never use Autopilot there. I'll switch it off, drive around this corner, and turn it back on."* (R038)

### 3.2.6.3. Operational
**Longitudinal control**

This sub-theme addressed driver-initiated disengagements caused by FSD Beta's control over braking and acceleration, which often did not meet drivers' expectations in terms of comfort and safety. Drivers reported disengaging the system to exceed the speed limit for reasons of safety and efficiency. Specifically, drivers reported increasing the operating speed to decrease the gap between the Tesla and the vehicle in front, overtake a slower vehicle, or decrease the operating speed of FSD Beta.

*"In scenarios where I need to go above the speed limit to get around a car safely, you have to disengage Autopilot because it will not do that for you. Autopilot currently does not have the ability of going above its speed in order to get out of a bad situation. It only knows how to slow down."* (R031)

---

[1] Interviewee shows researcher the situation on YouTube video documenting the situation.
[2] Interviewee shows researcher another situation on YouTube video.

*"I will disengage it when there's a long trail of cars and everybody else has got four feet or something in between each vehicle, I'll turn off Autopilot and then scooch the car forward, and then re-engage it so that the gap between me and the other cars is similar to what everybody else has."* (R037)

*"I did not trust the FSD Beta to brake in time, so quite often I would disengage the FSD Beta because I didn't feel that the vehicle was programmed to my safety standards. I just felt that it was just going too fast, too close to other vehicles when other vehicles were at a stop sign or a stoplight. So, I'd say 90% of the time I disengaged the FSD Beta for braking purposes."* (R041)

*"The only time is when sometimes cars slow down quickly, and it feels like it won't stop until the end. It's hard for you to tell because it stops later than you would stop, but you're nearing a point where you have to make a decision. Where do you let it? Do you? Do you wait, and see if it does its thing, or are you just gonna stop on your own? So, I stop on my own 80% of the time, and that's the only time, and I'm pretty sure it would stop, but that's the only part of whatever feels unsafe."* (R070)

*"Sometimes I don't even trust that it's gonna stop. 'Am I gonna stop?' I've had plenty of times where it was a controlled left turn, and the light was red, and I was going 55, and it changed lanes, went into the turn lane, and it was gonna go right through without stopping. That car was not gonna stop, that I was just gonna blow right through that red through the controlled stop, and I don't know what would happen."* (R076)

**Undesired lane changes**

This sub-theme addressed driver-initiated disengagements caused by the system performing unexpected, unnecessary, too aggressive, or too conservative lane changes.

*"FSD Beta – I can't look away even for a half second. It just changed lanes on its own as a motorcycle was rapidly approaching. I don't want to kill the motorcyclists. It's like 'Oh, I need to change lanes.' It randomly decided to*

*change lanes. Didn't give me time to do a head check because it just goes as soon as it wants to. Why are we changing lanes? It was unnecessary. It was unexpected. I did the head check, saw the motorcycle in my rearview mirror rapidly approaching, and took over. It likely would have been rear-ended, or the motorcycle would have had to take extreme evasive action to avoid it."* (R007)

*"So, it was like 'I'll change lanes', and I'll be thinking 'Why the heck are doing that for?' You know, 'I must say, I'm gonna right lane, I'm gonna make a right turn.' It'll go to the left lane for something, and then go back to the right lane again, and I'm thinking 'You know why I'm staying in the right lane?' You needed game theory going on in the left lane. Just turn it off because it will do things that maybe wasn't dangerous, but it just wasn't necessary.* (R072)

*"We disengage Navigate on Autopilot because sometimes it changes lanes too much, so then we'll turn it off. It will change lanes even while you're going down the freeway, look for the best lane, but sometimes that's annoying."* (R095)

*"It's trying to make a lane change and it's just freaking the other drivers out. Sometimes it'll try to make a lane change on the freeway, and there won't be quite enough space between people, so it'll just stay there, and stick there with blinker on, either holding up traffic or confusing the people behind me so then I feel unsafe, but then I just take control of the car."* (R098)

*"I was up in North Carolina on FSD Beta, and it was a two-lane divided highway going 50 miles an hour, and all of a sudden it said 'I wanna get in the left lane' so I thought 'Well, that's weird. There's nobody around. We don't need to be in the left lane', so I corrected it and made it go back to the right lane, and it did it again, and I corrected it again, and then this poor lady drives up in the lane beside me, and is trying to pass me in it, and it tries to do it again, and I said 'No, this Is not O.K.', and I turned it off. So."* (R100)

**Misidentifying correct lane**

This sub-theme addressed driver-initiated disengagements caused by FSD Beta driving into bike, bus, or parking lanes.

> *"It will almost always choose the wrong lane to be in. Sometimes it'll get in the parking lane. Sometimes it'll get in the correct lane. You never know. So, you really have to pay attention to that."* (R010)

> *"It does some weird things when it's coming across a bike lane. It mainly let me try going to the bike lane. Those things get better over time as people report them, but those are all that kind of things that are weird now."* (R034)

> *"They just need to program it so it can read the word 'bus lane'. In my commute, there's a bus lane, and you stay out of it, and that is on the right side of the road. It wants to go into that right lane, every time into the bus lane, and I have to cancel it."* (R081)

> *"I will disengage if it's not handling an area correctly. So, for example, the white bike lanes, the car will sometimes attempt to go there, and if I see it happening, I will disengage FSD, report it, because it shouldn't be doing that."* (R094)

**Creeping**

One of the main reasons for the driver-initiated disengagements at intersections was due to the creeping behavior of FSD Beta, which was causing confusion and annoyance to the drivers.

> *"If you make a right turn, and you need to get over multiple lanes immediately, the car will always do a right turn, and then take its sweet time to get over, and usually it seems it's going to be stuck getting, so I will just do it myself. It makes the narrowest right turn possible, and then slowly start stepping over."* (R050)

> *"On FSD Beta, I actually put it on the accelerator more often than not because it's just really slow at intersections. So, I manually have to override it with my* *foot to make it go through an intersection faster, or else I'll miss my chance, and then confuse other people in intersections."* (R060)

*"The big reason that I don't use it all the time is just there's all these social norms regarding how people drive, and people do not like it when you drive differently than they do. Even in the simplest of things where the car is pausing at a four way stop and then slowly, inexplicably creeping through the intersection when it's clearly time to go. People stop, and they look at you as you drive past, raise both hands. 'Go!' 'Hey, look, no hands!'"* (R068)

*"I'm sure you've heard this from other people about the creeping. So, when you're coming up to a stop sign, it does this thing called creeping. So, it does stop, and then it creeps forward slowly to make sure it's safe to go through the intersection. There is nobody in any direction, and it just sits there, and I just nudged the accelerator pedal, just tap it a little bit, and then it's like 'Ohh, ok', and then it'll finally go, and make the left turn."* (R069)

**Rolling through stop signs**

Some participants brought up the National Highway Traffic Administration's (NHTSA) efforts to regulate FSD Beta's practice of 'rolling through stop signs', which, although it represents an unlawful action, is a frequently observed human behavior in certain regions.

*"We're already dealing with the National Highway Traffic Safety Administration regulating Tesla where before Full Self-Driving, Beta would do rolling stops at stop signs and depending on where you live, that's a natural normal thing to do, but now Autopilot always has to stop at stop signs, and that makes the driving experience less natural."* (R038)

*"Another big reason is the NHTSA. They disabled the rolling stock functionalities of FSD Beta. It makes the car fully stop, which causes a huge traffic delay. If there's a clear intersection, almost everyone does rolling stops. It's becoming a little frustrating because the NHTSA stepped in, and now we have no rolling stops, which makes the car sit at a stop sign for a*

*significantly amount of time. So, I'm finding myself using the Beta less."* (R094)

*"At an intersection and there's lots of cars, a lot of times I'll just take over because it's too awkward. It hesitates too long to go, and I'm a very assertive driver. So, I get frustrated with other drivers who sit at stop signs for too long, like, 'Why are you waiting for the other person to get all the way through the intersection? There's no one else coming.'"* (R096)

**Running red lights / stop signs**
This sub-theme addressed driver- and system-initiated disengagements resulted from FSD Beta running a red light or a stop sign.

*"It actually has tried running a red light for me. I was following one of the test cars for one of the self-driving car companies, which ran a very late yellow, but then my car followed and when the light turned red, it wasn't slowing down. It wasn't reacting to it at all, so I took over."* (R003)

*"FSD basically tries to run this red light, and it basically is trying to take a ride on red. This is illegal. Illegal right turn. But it's trying to do, but I stopped because it can't see beyond that truck. There's no way that it knows it has the right of way."* (R011)

*"Then you would come up to the stop sign, you would slow down to stop, and it just kept going. It was just gonna go right through it, and that's when I hit the brake. As soon as you touch the brake, the whole Autopilot turns off. So, if I had not hit the brake, it would have gone right through that stop."* (R069)

*"It can run a stop sign. If the stop sign is not on the map to let the car know you have to stop here, the vision system cannot see and recognize with enough distance to give it enough time to come to a stop. I tested yesterday at 45, and in fact it ran the stop sign and then came to a stop afterwards on the other side."* (R096)

**Turning-on-red**

This sub-theme addressed driver-initiated disengagements in 'turning-on-red' situations. In turning-on red situations, cars are allowed to turn into the direction of traffic at a traffic light showing a right signal (Preusser et al., 1982). FSD Beta was sometimes not able to detect these traffic signs, resulting in drivers disengaging the system to avoid annoying other drivers or losing time while traveling.

> *"It still needs a lot of intervention. So, a typical example is drive up to a four-way-stop signs. Sometimes you have one or two cars around. So, we need to figure out who takes the turn, and typically with FSD Beta it doesn't actually do a very good job of going up at the full stop sign. So, it needs to go right away instead of waiting for other cars, but it'll always kind of wait for other cars, and so usually I like to press the accelerator pedal to kind of get to go."* (R003)

> *"Many of the intersections around me do not turn on red signs, and FSD Beta doesn't recognize those signs, so previously with FSD stopping at red lights, you had to manually intervene to make it proceed on green, and now I often have to manually intervene on red."* (R007)

> *"It doesn't know what 'turn on red' means. It doesn't know lane control signs. So, it is possible that if you're on those lanes, that FSD Beta might try to actually pass all the traffic and goes, 'Hey, this lane over here is pretty', but it's not supposed to be on it. I don't use it in this place because I'm pretty sure it's gonna do the wrong thing."* (R034)

> *"So those areas I tend to just maybe take over myself. Anytime there is a really weird sign, there's one right turn-on red sign in our area that a pedestrian can come up and if the pedestrian pushes the button, it'll give you a big red X over the crossbar of the traffic light. I don't know if Beta detects it yet, so I really am careful and ready to just hit the brake."* (R100)

**Unprotected left-turns**

Another unwanted FSD Beta behavior represents unprotective left turns. Such turns involve turning left at an intersection with a green light, instead of a green arrow, while oncoming traffic has the right of way. Human operators reported to disengage FSD Beta during unprotected left-hand turns.

> *"Waiting to turn unprotected left, there's no light there, and there's a really nice break in traffic, and it doesn't go, and I'm waiting for it to decide what it's going to do, and three cars come right toward me. I have my foot right over the brake pedal, and the car just suddenly tries to go. It just feels like it's jumping. These cars that are coming straight toward me, and a person wouldn't do that, ... unless they're trying to die."* (R018)

> *"It's an unprotected left turn, and the car will creep forward at the wrong time, when a car is just crossing, and that can be really scary because you're thinking, 'Oh, is the car just gonna run into this car? Why is it, why is it moving forward if there's traffic? So sometimes it does things that a human would never do, moving forward when it's obviously not time to move forward."* (R027)

> *"It was waiting to make a left-hand turn, and it started jittering, and moving forward a little bit when there was clearly traffic. A human would sit there, wait for traffic to be gone and go, whereas full self-driving realized, 'OK, I can move a little bit forward, I can move a little bit forward, and maybe there's a gap here.' There's no gap there. I wasn't willing to take that risk. The risk is if you're not paying close enough attention, it will probably crash the car."* (R045)

> *"There is a lot of times where the unprotected left turns, it just doesn't work. It's kind of scary, and you either have to press the accelerator because it just moving too slow. So, the car will kind of creep forward, and it will wait, and when the car passes by, it's still kind of thinking. 'Should I go? Should I stay right?', and then then it slowly starts moving."* (R058)

> *"The only times that FSB Beta make me feel unsafe is taking unprotected left turns mostly because there's no lights at all. Just knowing that there's vehicles *coming at you, and the vehicle's slowly creeping forward scares you because you don't know if it's just gonna gun it at the back at the worst possible moment. FSD Beta is moving way too forward, and ignoring the stopping white line. So, taking those left-hand turns are a little bit nerve wracking."* (R078)

**Other turning situations**

Human operators also indicated to disengage the automation in other turning situations, such as signalized left- and right-hand turning, and U-turning situations.

*"I think a week or two ago, I stopped at a stop sign and there's a highway in front. It was going to make a right turn onto the highway, but it stopped too far back from the sign. There are trees in the way. Instead of creeping forward, it decided that there were no cars coming and started to move out full speed into the highway, and there was another car coming, and I had to slam on the brakes because the other car might have run into me."* (R021)

*"Yesterday at a traffic light, it was gonna make a left turn. It was red, and green, cars were coming, and the wheel was moving like it was hesitating, thinking, and it was about to go. Two lanes of cars coming towards me. So, I had to jam on the brakes, and I hit the capture button like 'What on earth are you doing? You see everything else. You can't see these two lines of cars coming at you?' That would, that would have been an accident."* (R052)

*"Full-Self Driving Beta – it doesn't know how to make turns safely, even simple turns. Sometimes, it takes some weirdly fast. It'll stop. It will be like, 'OK, when it's time to go?', and I'll be like, 'OK, now!', and then it'll go really fast. It's like 'You could have just slowed down, and done that turn slowly.' So, it's not super safe."* (R059)

*"There's times it takes way too long for it to make a decision when making just a right turn, and there's times where it's an easy win. You can see there's no cars coming, no traffic, and it's still there taking side, scooting a little*

*more, soothing a little bit more. I'm like, 'Do you have to have confidence in this turn?'"* (R071)

*"On my normal test route, maybe it always gets in the wrong turn lane. It would get in it too early, and so I would just disengage, and get into the correct turn lane."* (R079)

### 3.3. Human operator's perception of other humans

Human operators may initiate the disengagement when they perceive that other humans, such as passengers and road users, become discomforted and frustrated by the actions of the automation's behavior, respectively.

#### 3.3.1. Passenger discomfort

Human operators reported to disengage Autopilot and FSD Beta as a result of passengers' discomfort and lack of trust in the car with standard Autopilot and FSD Beta engaged. Reasons for this included a perceived lack of control and the automation's erratic, harsh, and unpredictable behavior.

*"If my wife is in the car, I can't use it because she refuses to put up with it, even with disengaging, and she doesn't like all the beeping, and the alarms going off regularly."* (R007)

*"I know people who say 'Hey, I'm not driving Autopilot with my kids in the car', and my wife says, 'You're not using that thing with the kids in the car."* (R008)

*"I am less likely to use it when I have passengers in the car. I'd rather have that experience when I'm alone rather than doing it when I have a car full of family, and passengers who are like 'Oh, what's that thing been about? Why is it beeping at you?', or if I need to make a sudden correction, I don't wanna do that when I have got passengers that might be discomforted from that."* (R027)

*"It's my experience in my view and my feelings, and then that of my wife. I've had three occasions where had I not been absolutely paying attention and*

*being fully in control of the car, it would have caused an accident. My wife's comment is 'If I use the FSD Beta with her in the car again, then I should just go get the divorce papers.' It's that bad."* (R069)

*"My wife will not accept the FSD Beta being turned on when she is in the car with me because she does not trust it. She doesn't want me to turn it on, mostly because the way that it reacts to things can be very jerky and not very smooth. So, it's the harshness, and the non-warning of those situations that makes her very uncomfortable as a passenger and she won't use it herself when she's driving the car."* (R090)

### 3.3.2. (Anticipated) frustration

This sub-theme addressed driver-initiated disengagements to avoid confusion and rage of other road users interacting with Autopilot and FSD Beta on public roads.

*"The second biggest risk is road rage. Infuriating other road users from weird system behaviour. That computer does weird things that humans don't like, and it makes humans mad, right? It just makes them really angry, and then they are angry at you. That's actually been my biggest issue with the system. I'll just turn the system off. Anytime someone following me closely, I'm nervous, this system is going to break erroneously, and produce either an accident or anger."* (R002)

*"Drivers behind you are going to get annoyed and frustrated and that might cause them to behave in an anxious or unsafe way. That would put me in an unsafe condition, so I would intervene by either taking over the driving, or at least accelerating more than the car would."* (R012)

*"I really don't have an intervention unless somebody's behind me pushing me, and they get frustrated because it takes a turn too slow for them, and then I'll just disengage."* (R062)

*"Most of my disengagements are because I don't like how I perceive it will make other drivers feel because I'm a very courteous driver. When Autopilot*

*starts creeping out, I do feel it's going to move towards them, and that's going to make them feel I might dart out, and so how are they gonna react?"* (R085)

*"That indecisiveness about shifting to a lane and then shifting back when it corrects itself – it's not always the best for traffic behind you. They'll honk at you like 'Hey, what are you doing?' I don't wanna deal with that. Personally, I don't want people honking at me."* (R094)

### 3.3.3. Reckless behavior

Another reason for disengaging automation was the reckless behavior of other drivers, such as drivers following too closely from behind or swerving into the lane of the Tesla.

*"It was only at one time that I was on the Autopilot, and a car merged in so quickly that I actually had to come take control of the vehicle because my vehicle was not going to stop because he just merged out of nowhere."* (R020)

*"Yesterday, we had about four incidents where we had 18 wheelers cross over the line. So, it's something you want to keep an eye on to make sure that they're not going to come over in the lane."* (R024)

*"Again, somebody coming up behind you too quickly is they don't expect the car to be there. You understand that's a very complex situation coming up or that there are people driving around you way too aggressively who might be impatient, not wait for the car."* (R065)

*"I typically don't have to intervene most of the trip unless it's a situation that's more caused by another driver. They're driving erratically or there's something along those lines that I don't trust the system enough."* (R090)

*"One time when I was in a highway, I did see something a little weird. There was a truck with a really long flatbed that had nothing on it, and that truck tried to turn into my lane, and the car didn't realize that it was a really long trailer. So, I had to just slow it down, hit the brake."* (R100)

## 3.4. Automation's perception of human operator

The automation may initiate the disengagement itself if it perceives human operators becoming complacent. Both automation and human operators may initiate the disengagement of the automation with drivers exceeding the speed limit.

### 3.4.1. Complacency

Complacency of drivers constitutes a rationale for the automation to initiate its own deactivation.

> *"So, if you're paying attention, it's no problem. You know when to take over. If you're not, all of a sudden, that thing starts panicking. All it really does is basically slow down, and slam on the brakes, and stopped."* (R006)

> *"Autopilot never does disengage except for once. When I first dropped the car, and I was just not paying attention, and it threw me off because I wasn't paying attention."* (R066)

> *"There's been a few times I forget to do the nudging because it'll flash blue, and then it'll flash brighter, and it'll beeping at you, and a couple times, I was listening to music or like on a phone call or something, and then I may have forgotten to do that, and then it will get mad at you, and beep, and then it'll kick you out. It just tells you to take over, and then you can't use it until you park the car and start again. I think that's happened twice."* (R070)

### 3.4.2. Inappropriate amount of torque applied to steering wheel

Drivers applying an inappropriate amount of torque to the steering wheel resulted in unintentionally disengaging the automation.

> *"It doesn't take much to disengage it. Then you practically have taken over even though you didn't need to. So sometimes, I might hover my hand over the wheel. That way, I don't accidentally come disengage it during a turn."* (R032)

*"Full Self-driving Beta is really, really easy for taking over because you keep your hands on the steering wheel, and if it jerks the steering wheel with your hands on it, you're just gonna automatically disengage."* (R049)

*"They could get off the steering wheel nag because actually you could apply too much torque while you're trying to make sure it knows you are paying attention, and that takes it out of Full Self-driving Beta, and that could be a shock if you aren't ready for it."* (R062)

*"So, what happens is on a curve, it turns the steering wheel. So, your hands are on it, but if it wants you to nudge it on a turn, there's a chance you nudge it a little too much, and then you're on a turn with no Autopilot anymore."* (R070)

*"It's very hard to sometimes not accidentally disengage Autopilot when you're trying to show the car that I have my hands on the steering wheel."* (R088)

### 3.4.3. Speed rule violation

This sub-theme addressed driver- and system-initiated disengagements that occurred when drivers exceed the speed limit. Drivers reported initiating the disengagement to exceed the speed limit for reasons of safety, efficiency, or a lack of knowledge. The system disengaged when drivers exceeded the speed limit.

*"Now that I'm on the FSD Beta program, and I need to go just a little bit faster to get enough space between me and the car behind me to make a lane change, it will actually disengage, and put me in what they call Autopilot jail, so I cannot reengage Autopilot until I completely stopped and parked the car."* (R033)

*"Sometimes you'll go over the limit, and the car will start to scream at you. So, when that occurs, I have to manually turn it off. I can speed around people, and then I can turn it back on. Then I don't have to worry about getting kicked out for the remainder of the drive."* (R056)

> *"When I first bought the car there was nobody else around. I simply went too fast, and Autopilot kicked me off because I noticed that I went above 90 mph."* (R066)

> *"It used to kick off if you hit 80 miles an hour, and in Arizona, if you're doing 80 miles an hour, you're basically in people's way. So one time I was trying to get by a truck, and you get these flashing lights, and it kicks you off, and you're in jail until you stop, restart the car. You can't just kick it back in."* (R082)

> *"One main problem I have with Autopilot just disengaging is I wanna pass the truck, and I'll start accelerating around the truck and if I go over 82 miles an hour, it disengages, which is frustrating."* (R095)

### 3.5. Automation incapability in environment

This section covers the disengagements initiated by both the automation and human drivers in environments exceeding the perceived capability of the automation.

#### 3.5.1. Weather

This theme addressed driver- and system-initiated disengagements in inclement weather conditions, with respondents disengaging the system and the system asking drivers to take back control in poor visibility conditions given the current limitations of the sensor suite.

> *"I think the best example would be in bad weather because Autopilot heavily relies on all the cameras, and we have very bad weather here in Canada sometimes. When the cameras gonna cover it up too much with the snow, it will sometimes abruptly disable itself, and it does the warning beeps and everything, and that's a little jarring. I've never had it really lose control because it's slippery roads, but it's always in the back of my mind in bad weather."* (R016)

> *"It did disengage one other time. Last winter, there was like a slush on the freeway, and it was all stuck to the front of my car where the radar is, so*

*Autopilot wouldn't work because all the slush was covering the radar."* (R043)

*"Whenever the road isn't visible due to poor weather, I feel much more unsafe in that situation than I would if I was driving myself manually, so probably one of the only instances where I can say that I would much rather be driving than to allow Autopilot or Full Self-Driving to be active."* (R065)

*"In really bad weather, Autopilot is going to start chirping and making noises to tell you to take over when water gets on all the cameras, and it can't see any longer. So, normally if it starts doing that more than once or twice, I'll just turn it off until I'm done driving through the rainstorm."* (R070)

*"Most of the times that it would disengage would be weather-related. So, there's certain times where you're getting rain or snow, and it detects that you can't engage FSD Beta, it'll tell you there's inclement weather, you can't use it."* (R087)

### 3.5.2. Non-standard roads

This theme addressed driver- and system-initiated disengagements due to inconsistent lane markings and differences in lane width, which often resulted in the car re-centering itself within the lane.

*"I traveled across the U.S., and there were few states where they would have a dotted line for the lane line. If there's an exit lane with no dotted line, the car thinks the lane is very wide, and it's always trying to center itself between what it thinks is the lane, so it might center itself between the old lane and the new exit lane, and then it comes to the separation point, and makes some last second crazy decision. That's definitely scary. I'll just turn the system off."* (R007)

*"If you encounter an area where the paint disappears, the car tries to center itself in. Now if it's a two-lane highway, it may try to center itself, which is nerve wracking because part of me is not sure what would happen if there's a*

*car near me. There have been times where I've actually turned it off to get past those areas. Then I'll turn it back on."* (R056)

*"Earlier when I came to an on-ramp, off-ramp where there was no line, it could not see the line properly and then it disengaged, and the alarm went on, and it started slowing down. So, I took over."* (R073)

*"As soon as it makes the turn onto a road that has no markings, all that safety feeling goes out the window because it doesn't really know how to deal with unmarked roads well yet. So, it will try to stay in either the center of the road, and then it'll go to the right of the road, and then it'll go very slowly because it doesn't know if the edge of the road is really the edge of the road. So, I'll usually have to take over for that situation."* (R074)

*"I'm not sure that the car is gonna be able to read the lanes correctly because they're painted out, scratched out. You know you don't have clearly defined lines, and in most cases the Full Self-Driving and Autopilot disengages themselves in those situations anyway."* (R102)

### 3.5.3. Curves, hills

This sub-theme addressed driver-and system-initiated disengagements in curves (sharp turns) and on hills.

*"The largest issues I've historically had with Autopilot is, it doesn't turn sharply enough on sharp turns, so I get quite a bit of anxiety justifiably going into sharper turns. I'm always especially paying attention, and usually would disengage on sharper turns just because it wouldn't be steering enough so my steering would disengage Autopilot to keep it in the lane."* (R007)

*"There is one specific scenario that was so scary. Here[3] it'll get into the turn lane, and then it suddenly lurches to the right. It's doing some sort of curve optimization for smoothness, for driver comfort but it ends up leaving the lane, and this why whenever new Beta drivers are coming in, I always tell them*

---

[3] Interviewee shows researcher situation on YouTube video.

> *'You have to keep your hands right here because it will turn, and if you have not turned it back, it would have crashed into something, right?'"* (R011)

> *"There are certain sections of freeway where there are tight turns when everyone is going really fast, and I'll just manually take over on the really hard turns."* (R042)

> *"It struggles a little bit with tight turns and creeping. There's a balance, you want it to learn, but you don't wanna annoy people and you don't want to have to take over if you don't have to."* (R100)

### 3.5.4. Object detection

This sub-theme addressed driver- and system-initiated disengagements in response to stationary and non-stationary objects and events in the environment. Drivers decided to disengage the system in these situations because the system failed to detect and maneuver around these objects, or because of the system's lane centering, which resulted in an uncomfortable close proximity to these objects.

Stationary objects and events in the environment included potholes, parked cars, road debris, construction, cones, bushes/trees, buildings, gates, and railway crossings.

> *"I'm always nervous going hit a curb. It's going to run them over. I've run over many pieces of trash and things like that and potholes. Yes, if there are potholes, a lot of times I'll turn it off, and go around a pothole because this system doesn't recognize potholes."* (R002)

> *"Here[4] is one of those disengagements that is dangerous, and you need to intervene. So, there's an object in the road I didn't see until my fiancé said 'Hey, there's something in the road'. I had to disengage, and you can see that there's nothing visual as here. So, I would have hit that if I didn't take over."* (R011)

---
[4] Respondent showed researcher a YouTube video documenting the specific situation.

> *"I've been on Full Self-Driving Beta since October of last year, and I was going to go to the grocery store, and my car drove me there, and that's a nice experience all the way into the parking lot where it stopped, and then drove me home, and everything was fine until it almost drove into a building. I had to grab the wheel, put on the brakes."* (R018)

> *"I've had moments where even going out of my own driveway, I engaged Beta, and it immediately started making a sharp left into a tree by the side of my driveway, and if I had not intervened, it probably would have hit the tree. My driveway is straight. It doesn't get any simpler than that, so something was missing there."* (R068)

> *"That same railroad crossing I was approaching, and there was a train going through. This is rural. There's no gates that come down, there's no lights that flash. There's no bells that go off. It's just a stop sign. I approached the stop manually, and while I was stopped, I decided to turn Beta, and what it was displaying on the screen was a line of semi-trucks, and then the visualization disappeared, and it started to accelerate, and there's a train creeping forward. It was just like full normal acceleration. I'm not gonna let it get very close to a moving train. So, I hit the brakes as soon as it started to accelerate at normal velocity."* (R096)

Drivers reported disengaging the automation in situations with non-stationary objects and events in the environment, such as other vehicles in adjacent lanes, emergency vehicles, and vulnerable road users.

> *"It tried to hit a woman in the crosswalk yesterday. She's waiting for the walk indicator so she can cross, and I'm waiting to turn right. I'm on full-self driving, and I'm looking at her. She's just waiting for the light. I think 'My car's gonna try to hit her I know. I was gonna try to hit her I know.' The light turns green. Ohh. Turned to hit her."* (R018)

> *"I do not trust it around pedestrians and small children and dog walkers because it's too close to a human. That Tesla will approach the pedestrian and giving them a very small birth, and they would be like 'What are you* *doing? There's plenty of room.' It's not quite human so I don't always trust it around humans because it's not as courteous as a human would be, right?'"* (R031)

*"It is not that great at picking up on emergency vehicles yet, so if I see a police officer off the distance with its flashing lights, currently on the Autopilot version I have, it will start to slow down usually way too aggressively. Other people around me are not slowing down yet. I usually have to take over in those situations."* (R074)

*"FSD Beta – whenever there are many, many turns with lots of pedestrians, I'm not going to. It's too scary. I don't want it to hit anyone. I have to disable it because there's no point in risking someone's life or my freedom in order to test out something that wouldn't be able to do the job."* (R074)

### 3.5.5. Intersections

This sub-theme addressed driver- and system-initiated disengagements in intersections, including roundabouts. Respondents reported disengaging the system due to its limitations in handling the situation at intersections safely or their fear of getting rear-ended due to unexpected or unnatural system behavior. Respondents also reported to disengage FSD Beta at intersections due to the system's lack of negotiation and interaction skills with other road users.

*"When you're at an intersection, and there's four cars at this intersection, and they're all trying to encourage the other driver to go first. The car doesn't really recognize that. So, at that point, I will have to either take over, or I'll have to hope that the car goes pretty soon."* (R017)

*"Roundabouts, for example, are a very challenging scenario for FSD Beta still, and so it will almost always come to a complete stop at a roundabout or before entering the roundabout, and so I feel a sense of social anxiety if there's a car behind me, and the car just completely comes to a stop. I will tap on the accelerator to encourage the car to go through the roundabout."* (R050)

*"If I'm at an intersection, it can be too slow, and there's a lot of cars behind me, and I'm like 'Come on, no one's there. Come on.' Sometimes I think 'Accelerate because you can do that. You can accelerate.' It'll go and it'll do the turn. I don't always do that because certain situations, certain intersections are too big. So, then I'll just disengage, do the difficult spot, and re-engage."* (R052)

*"Especially things like roundabouts. It is just far too unsure. It's very jerky at times, it'll stop and start, stop, and start, and when you've got other traffic around you, I don't wanna deal with that. So, I'll typically turn it off. I typically only use it for highway driving."* (R056)

*"It's not a finished product. Always, every intersection I'm like, 'OK, we're gonna do this. Right. Great. Wonderful. Doing good. Next one. You can do this. Right. OK, good. Ohh. There's a little conservative here. Let me flag this.' Then the next time, I know, 'OK, it might be a little conservative here. It might get confused, and be prepared to take over.'"* (R057)

### 3.5.6. Discontinuities in road design

This sub-theme addressed driver- and system-initiated disengagements due to discontinuities in road design, such as in on- and off-ramp, lane merging, and splitting situations.

*"Common example is you'll be driving along, and a new lane opens up, and this week, the car just went into the left turn lane for no reason at all so I immediately intervened, and took it to the correct lane."* (R010)

*"I've had it disengaged near off ramps where it just doesn't understand that the lane is diverting. I haven't really had it disengaged based on anything else."* (R028)

*"When I took the exit on the freeway, when it just shouldn't, I didn't use it for the rest of the 30-minute drive. I just was like 'No, I am not using this because we almost just got in an accident.'"* (R043)

> *"It's got to adjust a few things, including exit ramps. It makes the maneuver way too aggressively, basically whipping you into that lane, and I'm not joking when I use the word 'whipping you' because you're there like 'Hey, lane change'. It literally throws you like this as it makes the maneuvering to the other lane. I have to disable it every time it needs to take that style of exit."*
> (R071)

### 3.5.7. Complex, heavy traffic

This sub-theme addressed driver-initiated disengagements in complex and heavy-traffic situations.

> *"So, if it was two in the morning, and there was no one else on the freeway, I would just kind of double check, and let it do its thing and not disengage, but usually there would be traffic around, so usually I'd need to disengage."*
> (R007)

> *"If there's a lot of cars on the road, I do not use Beta, and why is obviously because sometimes it tries to kill me."* (R023)

> *"If there's a lot of traffic, or if there's a lot of vehicles around, I probably won't use it because I don't want to intervene with the traffic. I don't wanna cause an unsafe situation, so I won't use it, but if it's empty, it's a dead area, then I'm gonna use it."* (R048)

> *"So, I only use it in areas where it's not gonna do anything bad, right? I use it in areas where I would feel comfortable training a brand-new driver, and so I don't feel safe using it on busy streets."* (R059)

## 4. Discussion

This study presents the findings of semi-structured interviews conducted with 103 users of standard Autopilot and FSD Beta, focusing on factors contributing to driver- and system-initiated disengagements of Autopilot and FSD Beta. Analysis of the interview data led to the extraction of main categories and sub-categories representing factors of disengagements of Autopilot and FSD Beta.

## 4.1. Triadic model of automation disengagement

The categories informed the development of a triadic model of automation disengagement, which treats human operators and the automation as equivalent agents. Each agent perceives the performance and capability of the other, initiating disengagement accordingly. The idea that both human operators and the automation can initiate the disengagement is consistent with existing literature (Endsley, 2019; Favarò et al., 2018). Human operators may initiate the disengagement if they perceive the automation's capabilities to be lower than their own in handling a maneuver safely. The triadic model of automation disengagements considers not only the impact of automation behavior and environmental characteristics, such as road infrastructure and design, on human operators but also on other humans (e.g., passengers, other road users). Previous driver behavior and automation models primarily featured the driver and automation as main actors (Banks et al., 2014; Michon, 1985; Rasmussen, 1983). Our model resonates with the principle of distributed cognition, which implies that information does not only exist in the minds of individuals but also in the objects and artefacts that individuals use as well as in the interactions between individuals and objects and artefacts (Stanton et al., 2010). As road traffic increasingly incorporates sensors, computers, and communication systems, the concept of distributed cognition is expected to become even more important in understanding and managing interactions within the transportation system.

The main reasons contributing to disengagement of the automation pertain to human operator's relevant states, human operator's perception of the automation, and human operator's perception of other humans, as well as the automation's perception of the human operator.

## 4.2.    Human operator's relevant states

Our research demonstrates that human operators opt to disengage automation due to both permanent and transient states. A permanent reason for disengaging automation involves the enjoyment of manual driving. Negative transient states include frustration, stress, and embarrassment. Prior research has identified trust, mental workload, and situational awareness as factors impacting automation disengagement (Dzindolet et al., 1999; Favarò et al., 2018; Parasuraman & Riley, 1997; Reagan et al., 2021; Wilson et al., 2020). Our study introduces additional psychological states (e.g., frustration, stress, embarrassment) that influence automation disengagements.

## 4.3.    Human operator's perception of automation

Human operators and automation may initiate disengagement in anticipation of automation failure in situations exceeding the perceived capability of the automation. Studies have shown that drivers frequently disengage automation in an anticipatory or precautionary manner (Endsley, 2019; Favarò et al., 2018;

Gershon et al., 2021; Lv et al., 2017; Morando et al., 2021; Mueller et al., 2021). Some reported driver-initiated disengagements might have been unnecessary, as it remains unclear whether the system would have handled the situation safely without driver intervention. Consistent with Gershon et al. (2021), drivers indicated initiating disengagement not to mitigate immediate risk but to perform tactical maneuvers exceeding the automation's capabilities (e.g., passing a vehicle in front).

The interviews also showed that drivers may opt to disengage automation when it demonstrates unnatural, non-human, or undesirable behavior at the operational, strategic, and tactical levels of decision-making (Michon, 1985), such as harsh deceleration (or phantom braking), erratic steering wheel movements, steering into adjacent traffic, choosing the incorrect route, unanticipated lane changes, misidentifying the correct lane, or creeping at intersections. While the phantom braking behavior associated with assisted and partially automated driving represents a known vehicle behavior (Nassi et al., 2020), our study offers new insights into additional vehicle behaviors (e.g., steering into oncoming traffic, undesired lane changes). Some of these behaviors are safety-critical and could have resulted in an accident if the driver had not intervened. Potentially safety-critical system behaviors (e.g., steering into oncoming traffic) could be mitigated through improvements in automated driving design (see Morando et al., 2021) or driver monitoring systems. These observations relate to the operational design domain (ODD) concept, which describes the conditions under which automated vehicles can function (SAE International, 2021). Our study supports existing research on external and environmental factors contributing to automation disengagement, as well as the need for human operator intervention in challenging situations (Boggs et al., 2020; Favarò et al., 2018; Lv et al., 2017).

### 4.4. Human operator's perception of other humans

Our study emphasizes the significant role other humans play in shaping disengagements of automation. Human operators may choose to disengage automation if passengers and other road users feel uncomfortable or are angry due to negative automation behavior. Consequently, the human operator possesses the 'theory of mind ability' (Baron-Cohen, 1995; Leslie, 2001) to infer how other humans interacting with automation feel about its behavior. The perceived risk of accidents and loss of control may be higher for passengers than for the driver (Häuslschmid et al., 2017), and other road users who lack direct control over the actions of the automation. Our research suggests that if other humans are uncomfortable with drivers engaging the system, the anticipated safety and efficiency benefits of road vehicle automation may not be fully realized. For instance, drivers might opt to use the system alone without passengers onboard, testing and experiencing its limits, which may lead to increased single-person vehicle miles traveled (see Nordhoff et al., 2023). Car manufacturers should consider the impact of automation on passengers and other road users, designing the human-machine interaction in such a way that perceived safety and trust of humans inside and outside automated vehicles is promoted, and frustration and stress reduced.

## 4.5. Automation's perception of the human operator

Our study also revealed that automation decides to initiate disengagement if it, based on the hands-on-wheel sensor, detects that the human operator becomes complacent or violates traffic rules. The risk of complacency associated with partial automation is well-known (Banks et al., 2018; Reagan et al., 2021; Wilson et al., 2020). Analogous to the concepts of 'trust in automation' and 'trust in self', the automation should have accurate trust in humans and trust in self, initiating disengagement when the perceived reliability of itself is lower than the perceived reliability of human control. Driver education and training could play a crucial role in raising awareness of system limitations and improving system handling (e.g., the required amount of torque applied to the steering wheel.

## 4.6. Limitations and implications for future research

The data on driver- and system-initiated disengagements reflects the subjective perceptions of respondents. While disengagements are considered a safety risk (Favarò et al., 2018), it is not clear to what extent disengaging automation contribute to a decrease or increase in accident risk compared to not disengaging the automation. Respondents represented early adopters and were thus not representative of the broader population of users of partially automated cars. We recommend future research to perform studies in naturalistic driving conditions to investigate to what extent disengagements influence accident risk with a representative population of users of partially automated cars.

## 5. Conclusion

The study presents the results of semi-structured interviews with 103 users of Tesla's Full-Self-Driving (FSD) Beta system and standard Autopilot, exploring the factors contributing to driver- and system-initiated disengagements of both systems. The study proposes a novel triadic model of automation driver- and system-initiated disengagements, which takes into account the impact of automation behavior not only on the human operators of automation but also on passengers and other road users. The results have shown that the contributing factors leading to driver- and system-initiated disengagement of Autopilot and FSD Beta pertain to permanent and transient human operator states, safety-critical system behaviors (e.g., steering into oncoming traffic), other road users' behavior (e.g., road rage, reckless driving behavior), and road infrastructure and design factors (e.g., missing lane markings). The findings provide new insights into the factors contributing to disengaging partial automation, with valuable information on potential edge cases of automated vehicle technology. Our paper emphasizes that automation disengagement is not solely a human operator-based concern but also a social phenomenon, in which the operator may feel a sense of embarrassment or self-consciousness about the impact of the automation on others. This social dimension of distrust highlights the importance of considering the experiences and perceptions of other road users in addition to the human operator when evaluating the effectiveness and usability of automation systems.

## 6. Acknowledgments

## 8. Appendix

*Table 1.* Overview of questionnaire

| Question number | Question |
|---|---|
| Q1 | Do you have the Full Self-Driving Beta (FSD Beta) feature? (1 = Yes, 2 = No) |
| Q2 | Before the first time of using Autopilot and FSD Beta, did you watch / read / listen to information on how to use it? (1 = Yes, 2 = No) |
| Q3 | Please mention the type of information you consulted on how to use Autopilot and FSD Beta (website of Tesla (www.tesla.com), car dealer / sales point, online communities and forums, YouTube videos, newspapers and magazines, friends, family, colleagues, driver manual) |
| Q4 | Please describe your experience with using Autopilot and FSD Beta and the benefits and risks associated with using it. Please explain your answer. |
| Q5 | Have your expectations of using Autopilot and FSD Beta been fulfilled? Why / why not? |
| Q6 | Why do you use Autopilot and FSD Beta? |
| Q7 | Did you ever stop using Autopilot and FSD Beta (for prolonged periods of time)? |
| **Next, we would like to explore your perceptions regarding four general statements about the operation of Autopilot and FSD Beta.** | |
| Q8 | The current Autopilot does make driving autonomous. Is that correct? (1 = Yes, 2 = No, 3 = I don't know) |
| Q9 | There are no safety issues with Autopilot. Is that correct? (1 = Yes, 2 = No, 3 = I don't know) |
| Q10 | Autopilot is a hands-on feature. Is that correct? (1 = Yes, 2 = No, 3 = I don't know) |
| Q11 | Tesla FSD Beta is safer than a human. Is that correct? (1 = Yes, 2 = No, 3 = I don't know) |
| **With the next section, we would like to explore your perceptions of safety while using Autopilot and FSD Beta.** | |
| Q12 | Do you feel safe when Autopilot and FSD Beta is active? Why / why not? |
| Q13 | What / how do you feel when you feel safe / unsafe? Please explain. |
| Q14 | What is it about Autopilot and FSD Beta that is safe / unsafe? Please explain. |
| Q15 | Now please remember the situation / s in which you typically feel unsafe when Autopilot and FSD Beta is active and describe these situations. |
| Q16 | What can Autopilot and FSD Beta do to support your safety in Autopilot and FSD Beta? Please explain. |
| Q17 | Does feeling safe / feeling unsafe impact how you use Autopilot and FSD Beta on your next drives / in the future? Please explain. |
| Q18 | Has your perceived safety changed over time? If so, how? |
| **With the next section, we would like to explore your trust in Autopilot and FSD Beta.** | |
| Q19 | How would you position your level of trust in Autopilot and FSD Beta. (1 = I don't trust it at all, 2 = I don't trust it, 3 = I neither don't trust it at all nor trust it a lot, 4 = I trust it, 5 = I trust it a lot) |
| Q20 | What can Autopilot and FSD Beta do to support your trust in Autopilot and FSD Beta? |
| Q21 | Does your trust / distrust in Autopilot and FSD Beta impact how you use Autopilot and FSD Beta on your next drives / in the future? Please explain. |
| Q22 | Has your trust changed over time? If so, how? |
| Q23 | When you do compare yourself with other drivers, Autopilot, and FSD Beta, do you think you are ... (1 = A much worse driver, 2 = A worse driver, 3 = Not a better nor a worse driver, 4 = A better driver, 5 = A much better driver) (De Craen, 2010) |

| | **With the next section, we would like to explore how you typically use Autopilot and FSD Beta.** |
|---|---|
| Q24 | How do you typically place your hands on the steering wheel when Autopilot and FSD Beta is active? Please select the image that serves as best representation of your placement of your hands on the steering wheel when Autopilot / FSD Beta is active and explain your answer.<br><br>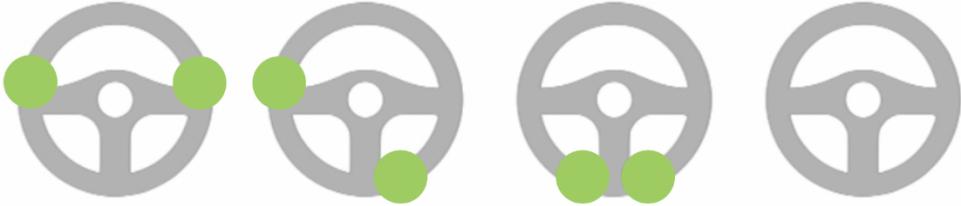<br><br>Figure is from Morando et al. (2021) |
| Q25 | Do you typically keep your hands on the steering wheel at all times? |
| Q26 | Are you typically fully attentive and alert at all times? |
| Q27 | How often do you typically engage in other secondary activities while Autopilot and FSD Beta is active? (Never, rarely, occasionally, frequently, always; monitoring the road ahead, talking to fellow travelers, observing the landscape, using the phone for music selection, using the phone for navigation, using the phone for calls, eating and drinking, using the phone for texting, watching videos / TV shows, sleeping) |
| Q28 | Do you disengage Autopilot and FSD Beta? Why / why not? |
| Q29 | Does Autopilot and FSD Beta disengage? When / in which situations? |
| Q30 | How do you typically place your eyes when Autopilot and FSD Beta is active? |
| Q31 | Do you typically keep your eyes on the road at all times? |
| Q32 | Do you typically monitor the vehicle and its surroundings at all times? |
| Q33 | How do you typically place your feet when Autopilot and FSD Beta is active? |
| Q34 | Do you typically stay prepared to take corrective actions at all times? |
| Q35 | Has your use of Autopilot (in terms of how you placed your hands on the steering wheel, eyes on the road, and feet) changed over time? If so, how? |